\documentclass[12pt]{amsart}
\usepackage{amsbsy,amssymb,amsmath,amsthm,amscd,amsfonts,latexsym,amstext,delarray,
amsmath,graphicx} 
\usepackage[margin=1in]{geometry}
\usepackage{color}
\input xypic

\usepackage[all]{xy}
\usepackage{setspace}

\usepackage[T1]{fontenc}
\usepackage[utf8]{luainputenc}
\usepackage{float}

\newtheorem{thm}{Theorem}[section]
\newtheorem{prop}[thm]{Proposition}
\newtheorem{cor}[thm]{Corollary}

\newtheorem{defn}[thm]{Definition}
\newtheorem{rem}[thm]{Remark}
\newtheorem{ex}[thm]{Example}

\numberwithin{equation}{section}

\def\cF{{\mathcal F}}
\def\cG{{\mathcal G}}
\def\cH{{\mathcal H}}
\def\cI{{\mathcal I}}

\def\cL{{\mathcal L}}

\def\cV{{\mathcal V}}
\def\cW{{\mathcal W}}

\def\val{{\rm val}}

\title{Graph Grammars, Insertion Lie Algebras, and Quantum Field Theory}
\author{Matilde Marcolli and Alexander Port}
\address{Mathematics Department, Caltech, 1200 E. California Blvd. Pasadena, CA 91125, USA}
\email{matilde@caltech.edu}
\email{aport@caltech.edu}
\date{}

\begin{document}
\maketitle

\begin{abstract}
Graph grammars extend the theory of formal languages in order to model distributed 
parallelism in theoretical computer science. We show here that to certain classes
of context-free and context-sensitive graph grammars one can associate a Lie
algebra, whose structure is reminiscent of the insertion Lie algebras of quantum
field theory. We also show that the Feynman graphs of quantum field theories
are graph languages generated by a theory dependent graph grammar.
\end{abstract}

\section{Introduction}

Graph Languages and Graph Grammars were introduced in theoretical 
computer science as an extension of the theory of formal
languages (linear languages), in order to model various types of parallelism in computation,
\cite{Nagl}, \cite{HEPT}, \cite{EKR}, \cite{Roz}. Instead of replacing nonterminal symbols with combinations of nonterminals and terminals in a linear string of characters,
the production rules of graph grammars replace a subgraph of a graph
with a new graph. The latter is obtained either by gluing along a common subgraph, or by
first performing an excision of a subgraph and then replacing it with a new graph. 
An analog of the Chomsky hierarchy of grammars exists for Graph Languages, see \cite{Nagl}. 
In particular, the context-free Graph Grammars are those where the left-hand-side of the
production rules is always a single vertex. Namely, no ``context" in the graph is taken into consideration
in deciding when a production rule can be applied: it just applies to any vertex. In this context-free
case the production rules then consist of inserting a new graph at a vertex of another graph.
This operation is reminiscent of the insertion operation that defines the Lie algebras of
Feynman graphs in the algebraic approach to renormalization in quantum field
theory pioneered in \cite{CoKr} (see also \cite{CoKr2}, \cite{CoMa}, \cite{EFBP}). In this paper we
show that, indeed, to certain classes of Graph Grammars (both context-free and context-sensitive)
it is possible to associate a Lie algebra, obtained by constructing a pre-Lie insertion operator
using the production rules of the grammar. We also show that the Feynman graphs of a given
quantum field theory are a graph language in the sense of the theory of formal languages.
This provides a new class of examples of graph languages, in addition to those arising in the
context of computer science (such as FFT networks, Petri nets, distributed parallelism), see 
the articles in \cite{EKR} for several examples. Relations between the formalism
of algebraic renormalization in quantum field theory and aspects of the theory of
computation in theoretical computer science have already been investigated in \cite{Man},
see also the formulation of Dyson--Schwinger equations in the Hopf algebra of
flow charts in \cite{DelMar}. it would be interesting to see if a 
theory of Dyson--Schwinger equations can be formulated 
for Graph Languages, using the Lie theoretic approach of \cite{Foissy}.

\subsection{The Insertion Lie algebra of Quantum Field Theory}

In perturbative quantum field theory, one computes expectation values
as a formal series of Feynman amplitudes labeled by Feynman graphs.
These graphs are finite and the allowed valencies are constrained to match 
the exponents in the interaction monomial in the Lagrangian of the field theory. 
Graphs have a number of internal edges (connecting pairs of vertices)
and external edges (half edges).
The corresponding Feynman amplitude is a finite dimensional integral over a space of
momenta flowing through the graph, with assigned external momenta carried
by the external edges, and with conservation laws at the vertices.
These Feynman integrals are typically divergent, which leads to
the crucial problem of renormalization. The goal of a renormalization procedure 
is a consistent extraction of finite values from all these integrals that takes into account the
combinatorics of how divergent subgraphs are nested inside larger
graphs. Since the work of Kreimer \cite{Kr} and Connes--Kreimer \cite{CoKr},
it has become clear that the renormalization procedure can be formulated
algebraically in terms of a Hopf algebra of Feynman graphs. The 
algebraic Feynman rules are seen as algebra homomorphisms to a target
commutative algebra determined by a choice of regularization procedure, and endowed 
with a ``pole-subtraction" operation (Rota--Baxter algebra). See \S 1 of \cite{CoMa}
for an overview.

\smallskip

The Hopf algebra $\cH$ of Feynman graphs is a graded connected commutative
Hopf algebra generated by the 1PI Feynman graphs of the given quantum
field theory. The 1PI (one-particle irreducible) condition means that the graphs
are connected and cannot be disconnected by removal of a single edge.
A standard argument in quantum field theory reduces the combinatorics of
Feynman graphs to the connected case, and further to the 1PI case, see \cite{ItZu}.
The coproduct in the Hopf algebra is not co-commutative. It is given by
\begin{equation}\label{CKcoprod}
\Delta(G)=G \otimes 1 + 1 \otimes G + \sum \gamma \otimes  G/\gamma,
\end{equation}
where the sum is over all the (not necessarily connected) subgraphs $\gamma \subset G$,
such that the quotient graph $G /\gamma$ (obtained by shrinking each component of $\gamma$
to a vertex) is a 1PI Feynman graph of the theory. The Hopf algebra is 
dual to a pro-unipotent affine group scheme that is entirely determined by its Lie algebra. 
Connes and Kreimer gave a very explicit geometric description of this {\em insertion Lie
algebra} \cite{CoKr2} (see also \S 1 of \cite{CoMa}). 
On the vector space spanned by all 1PI Feynman graphs of the theory, one can define a Lie bracket
by setting 
\begin{equation}\label{QFTLiebracket}
[G_{1},G_{2}]=\sum_{v}G_{1}\circ_{v} G_{2}-\sum_{v'} G_{2}\circ_{v'} G_{1},
\end{equation}
where the sums are over all vertices in $G_{1}$ and $G_{2}$.
The expression $G_{1}\circ_{v}G_{2}$ denotes the graph resulting from
the insertion of $G_{2}$ at the vertex $v$ of $G_{1}$.
One can insert one graph into another by matching external edges to
the edges incident at the vertex. In the sum one counts all the possible
inequivalent ways in which the graph can be inserted at the given
vertex. This bracket indeed satisfies the Jacobi
identity and defines a Lie algebra, which can be related to the 
primitive elements in the dual Hopf algebra of Feynman graphs \cite{CoKr2} 
(see also \S 1 of \cite{CoMa}). A detailed survey of the use of Lie algebra methods
in Quantum Field Theory can be seen in \cite{EFBP}. The language
of Lie algebras in Quantum Field Theory provides an elegant formulation
of the Dyson--Schwinger equations (the quantum equations of motion
of the theory), and a general method for solving them in the Lie algebra
of Feynman graphs \cite{Foissy}.

\section{Graph Grammars and Lie algebras}

\subsection{Two descriptions of graphs}

It is convenient to consider two slightly different ways of assigning the data of a 
finite graph. The first is the one most commonly used in Combinatorics, while
the second is more frequently used in Physics.

\medskip

\subsubsection{Version 1:} A graph $G$ consists of a set of vertices $V(G)$ and a set of edges $E(G)$
together with a boundary map $\partial: E(G)\to V(G)\times V(G)$ assigning
to an edge $e\in E(G)$ its (unordered) pair of boundary
vertices $\partial(e)=\{ v_1, v_2 \}$. The graph can have looping edges if we
allow $v_1=v_2$ and it can have multiple parallel edges if $\partial^{-1}(v_1,v_2)$
can consist of more than one element. If the graph $G$ is oriented (directed)
then the boundary map consists of two maps (source and target)
$s,t: E(G)\to V(G)$. A system of vertex and edge labeling consists of two
sets $\Sigma_V$, $\Sigma_E$ of vertex and edge labels, respectively, and
functions $L_{V,G}: V(G) \to \Sigma_V$ and $L_{E,G}: E(G) \to \Sigma_E$.

\medskip

\subsubsection{Version 2:} 
A graph $G$ consists of a set $C(G)$ of {\em corollas} (vertices $v$ with valence
$\val(v)$ with $\val(v)$ half-edges attached to it) and an involution
$\cI: \cF(G)\to \cF(G)$ on the set $\cF(G)$ of all half-edges (flags) attached to
all the corollas. The set $E_{int}(G)$ of internal edges of $G$ corresponds to all the pairs
$(f,f')$ with $f\neq f'$ in $\cF(G)$ and with $f'=\cI(f)$. The set $E_{ext}(G)$ of external
(half)edges of $G$ consists of all the $f\in \cF(G)$ such that $\cI(f)=f$. A labeling system 
is given by a set $\Sigma_\cF$ of flag labels and a set $\Sigma_V$ of vertex
labels together with maps $L_{\cF,G}: \cF(G)\to \Sigma_\cF$ and $L_{V,G}: C(G)\to \Sigma_V$
with $L_{\cF,G}\circ \cI= L_{\cF,G}$.

\medskip
\subsection{Insertion Graph Grammars}

Using the first description of graphs, we define an Insertion Graph Grammar  as follows.

\begin{defn}\label{IGG1}
An Insertion Graph Grammar consists of data
$$ (N_E, N_V, T_E, T_V, P, G_S) $$
where the set of edge labels of graphs is $\Sigma_E=N_E \cup T_E$,
with $N_E$ the nonterminal symbols and $T_E$ the terminal symbols,
and the set of vertex labels is given by $\Sigma_V = N_V\cup T_V$, 
with non-terminal and terminal symbols given respectively by
$N_V$ and $T_V$. The start graph is $G_S$ and $P$ is a finite set of 
production rules of the form $P =(G_L, H, G_R)$, with $G_L$ and
$G_R$ labelled graphs (respectively, the left-hand-side and the right-hand-side
of the production) and with 
$H$ a labelled graph with isomorphisms 
$$ \phi_L: H \stackrel{\simeq}{\to} \phi_L(H)\subset G_L, \ \ \  
\phi_R: H \stackrel{\simeq}{\to} \phi_R(H)\subset G_R. $$
The isomorphism $\phi_L$ should be label preserving.
The production rule $P =(G_L, H, G_R)$ searches for a copy of $G_L$
inside a given graph $G$ and glues in a copy of $G_R$ by
identifying them along the common subgraph $H$, with new
labels matching those of $\phi_R(H)$.
\end{defn}

\medskip
\subsubsection{Context-free Graph Grammars}

We recall the notion of context-freeness for graph grammars from \cite{Nagl}.

\begin{defn}\label{contextfree}
An Insertion Graph Grammar as in Definition \ref{IGG1}
is {\em context-free} if $G_L =\{ v \}$ (hence $H=\{ v \}$ also).
It is {\em context-sensitive} if $G_L \neq \{ v \}$. In the context-sensitive case $G_L$
is called the {\em context} of the production rule. 
\end{defn}

A Chomsky hierarchy for graph grammars is described in \cite{Nagl}.

\medskip
\subsubsection{Insertion Graph Grammars and Flags} If we consider the second
version of the definition of graphs given above, we can formulate a slightly different
notion of Insertion Graph Grammars. For a subgraph $G'\subset G$ the set of 
external edges $E_{ext}(G';G)$ is defined as the union of the set $\cF_{G'}\cap E_{ext}(G)$
and the set of pairs $(f.f')\in E(G)$ such that only one half-edge in the pair belongs to $\cF_{G'}$
while the other belongs to $\cF_G\smallsetminus \cF_{G'}$.

\smallskip

In this setting, we describe an Insertion Graph Grammar as follows.

\begin{defn}\label{IGG2}
An Insertion Graph Grammar consists of data 
$(N_\cF, N_V, T_\cF, T_V, P, G_S)$, as in Definition \ref{IGG1}, with
$\Sigma_\cF = N_\cF \cup T_\cF$ the non-terminal and terminal
labels for flags. The production rules $P =(G_L, H, G_R)$ are as in Definition \ref{IGG1}, 
with the additional requirement that $\phi_L(E_{ext}(H,G_R))\subset E_{ext}(G_L,G)$
and $\phi_R(E_{ext}(H,G_L))\subset E_{ext}(G_R)$, where $G$ is any graph
the production rule is applied to, with $G_L\subset G$.
\end{defn}

\smallskip

The reason for this modified definition is that the notion of gluing of two
graphs $G_L \cup_H G_L$ along a common subgraph $H$ is formulated
by taking as set of corollas $$C_{G_L \cup_H G_L} =C_{G_L} \cup_{C_H} C_{G_R},$$
identifying the corollas around each vertex of $H$ in $G_L$ and $G_R$ and
then matching half-edges by the involution  $\cI(f)=f'$ with $f'=\cI_L(f)$ when 
both $f,f'\in \cF_{G_L}$, with $f\neq f'$, and $f'=\cI_R(f)$ when $f,f'\in \cF_{G_R}$, with $f\neq f'$.
If $\cI_1(f)=f$ and $f\in \phi_L(E_{ext}(H,G_R))\subset E_{ext}(G_L,G)$ with
$f=\phi_L(f')$, when $\cI(f)=f'$ and similarly for $\phi_R(E_{ext}(H,G_L))\subset E_{ext}(G_R)$.

\smallskip

In this setting, because vertices are always endowed with a corolla of half-edges,
we cannot state the context-free condition by requiring that $G_L=H=\{ v \}$. An
appropriate replacement of the context free condition is given by the following.

\begin{defn}\label{contextfree2}
An Insertion Graph Grammar as in Definition \ref{IGG2} is {\em context-free}
if $G_L=H=C(v)$, the corolla $C(v)$ of a vertex $v$, and all the vertices of graphs in
the graph language have the same valence. 
\end{defn}

In the case where graphs contain vertices of different
valences, these would still be context-sensitive graph grammars, with the
context specified by the valence of $C(v)$.

\medskip
\subsection{Insertion-elimination Graph Grammars}

We consider another variant of the definition of graph grammars,
where the production rules consist of {\em replacing} a subgraph by another one,
instead of gluing them along a subgraph. While the version discussed above
reflects the notion of graph grammars considered for instance in \cite{Nagl},
the version we discuss here reflects the use in other references (see for instance
\cite{Roz2}).

\smallskip

In order to formulate this version of graph grammars with the first notion of
graphs, we need to define the operation of removal of a subgraph from a graph.
Let $G'\subset G$ be a subgraph. Let 
\begin{equation}\label{EGGprime}
E_G(G')=\{ e\in E(G)\smallsetminus E(G')\,|\, \partial(e)\cap V(G')\neq \emptyset \}. 
\end{equation}
We define
$G\smallsetminus G'$ as the subgraph of $G$ with $V(G\smallsetminus G')=
V(G)\smallsetminus V(G')$ and with edges $E(G)\smallsetminus (E(G')\cup E_G(G'))$.
Thus, for example, removing a vertex $G'=\{ v \}$ means removing the vertex $v$ along 
with its star of edges. We then define Insertion-elimination Graph Grammars as follows.

\begin{defn}\label{IEGG1}
An Insertion-elimination Graph Grammar consists of data $$\cG=(N_E, N_V, T_E, T_V, P, G_S)$$
as in Definition \ref{IGG1}, where the production rule $P=(G_L,H,G_R)$ acts by
searching for a copy of $G_L$ in $G$, removing $G_L\smallsetminus H$
and replacing it with the graph $G_R$ glued along $H$.
\end{defn}

\smallskip

Using the second description of graphs, the removal of a subgraph $G'\subset G$ is defined by
cutting all edges in $E_{ext}(G',G)$ into pairs of half-edges, one attached to $G'$ and one to
$G\smallsetminus G'$. Thus, the set of corollas $C(G\smallsetminus G')$ is given by
the difference $C(G)\smallsetminus C(G')$ and the set of flags is given by
$\cF(G\smallsetminus G') = \cF(G)\smallsetminus \cF(G')$, with involution
$\cI_{G\smallsetminus G'}(f)=\cI_G(f)$ if both $f$ and $\cI_G(f)$ are in $\cF(G)\smallsetminus \cF(G')$
and $\cI_{G\smallsetminus G'}(f)=f$ for $f\in \cF(G)\smallsetminus \cF(G')$ with $\cI_G(f) \in \cF(G')$.
Notice that the two notions of removal of subgraphs differ in the way the edges connecting
a vertex of the subgraph to a vertex of the complement are treated: in the first case they are
removed, while in the second case a half-edge remains as an external edge of the
complement graph. We then have the following formulation.

\begin{defn}\label{IEGG2}
An Insertion-elimination Graph Grammar consists of data $$\cG=(N_\cF, N_V, T_\cF, T_V, P, G_S)$$
as in Definition \ref{IGG2}, with the requirement that 
$$  E_{ext}(\phi_L(H),G)=E_{ext}(\phi_L(H),G_L)= E_{ext}(\phi_R(H),G_R)=E_{ext}(G_R) . $$
The production rule $P(G_L,H,G_R)$ acts by searching for a copy of $G_L$ inside $G$,
removing $G_L\smallsetminus \phi_L(H)$ and replacing it with a copy of
$G_R\smallsetminus \phi_R(H)$, by matching the half-edges of $E_{ext}(G_R)$ to the
half-edges of $E_{ext}(G_L,G)$.
\end{defn}

\medskip
\subsection{Pre-Lie structures}

A (right) pre-Lie structure on a vector space $V$ is a bilinear map $$\triangleleft : V\otimes V \to V$$
satisfying the identity of associators under the exchange $y \leftrightarrow z$,
\begin{equation}\label{prelie}
(x \triangleleft y)\triangleleft z - x \triangleleft (y \triangleleft z) = (x \triangleleft z)\triangleleft y
 - x \triangleleft (z \triangleleft y), \ \ \ \forall x,y,z\in V.
\end{equation}

\medskip

A Lie algebra is a vector space $V$ endowed with a bilinear bracket $[\cdot ,\cdot]$
satisfying antisymmetry $[x,y]=-[y,x]$ and the Jacobi identity
\begin{equation}\label{jacobi}
 [x,[y,z]]+[z,[x,y]]+[y,[z,x]]=0, \ \ \  \forall x,y,z \in V.
\end{equation} 

\medskip

A pre-Lie structure determines a Lie algebra by setting
\begin{equation}\label{prelietolie}
[x,y] := x \triangleleft y - y \triangleleft x .
\end{equation}
The pre-Lie identity ensures that the Jacobi identity is satisfied.
A detailed survey of occurrences of pre-Lie algebras in geometry, physics, and
the theory of formal languages can be found in \cite{Burde}.

\medskip

One can obtain a group structure from a pre-Lie algebra structure 
(see \cite{AgraGam} and \cite{Manchon}) by considering formal series
$$ W(x)= x + \frac{1}{2} x \triangleleft x + \frac{1}{6} (x \triangleleft x) \triangleleft x + \cdots $$
with the multiplication operation
$$ W(x)\star W(y) =W(C(x,y)), $$
where $C(x,y)$ is the Baker--Campbell--Hausdorff formula
$$ C(x,y)=x+y+\frac{1}{2}[x,y]+\frac{1}{12} ( [x,[x,y]]+[y,[y,x]]) + \cdots $$

\medskip
\subsection{Lie algebras of context-free grammars of directed acyclic graphs}

Consider the case of a context-free Insertion Graph Grammar $\cG$ as in 
Definitions \ref{IGG1} and \ref{contextfree}, where the start graph $G_S$
is a single vertex and all the graphs $G$ are directed acyclic with a marked (root)
source vertex. The production rules are of the form $P(v,v_2,G_2)$, where
$v_2$ is the root vertex of $G_2$ and $v$ is a vertex of the graph $G_1$, to which
the rule is applied. The resulting graph 
$$ G_1\triangleleft_{v} G_2= P(v,v_2,G_2)(G_1) = G_1\cup_{v\equiv v_2} G_2 $$
obtained by applying the production rule to $G_1$ is also a directed acyclic graph
with root vertex the root $v_1$ of $G_1$. 

Let $\cV$ be the vector space spanned by the set $\cW_\cG$ all the graphs 
obtained by repeated application of production rules, starting with $G_S$.
The set $\cW_\cG$ is different from the graph language $\cL_\cG$, as it
also contains graphs whose vertices and edges are labelled by non-terminal
symbols. 

We then define the insertion operator $\triangleleft : \cV\otimes \cV \to \cV$ as 
\begin{equation}\label{insopCF}
G_1 \triangleleft G_2 = \sum_{v\in V(G_1)} P(v,v_2,G_2)(G_1) 
=\sum_{v\in V(G_1)} G_1\triangleleft_v G_2 .
\end{equation}

\begin{prop}\label{preLie1ins}
Given a context-free Insertion Graph Grammar $\cG$ as above,
the insertion operator \eqref{insopCF} defines a pre-Lie
structure on the vector space $\cV$.
\end{prop}

\proof We need to check that \eqref{prelie} is satisfied. We have
$$ (G_1\triangleleft G_2)\triangleleft G_3 =  \sum_{v\in V(G_1)}\,\, 
\sum_{v'\in V(G_1\triangleleft_v G_2)} \,\, (G_1\triangleleft_v G_2) \triangleleft_{v'} G_3 $$
where $v$ and $v'$ are glued, respectively, to the root source vertices 
$v_2$ and $v_3$ of $G_2$ and $G_3$. The choice of
$v'$ can be subdivided into the two cases where $v'$ is a vertex of $G_2$ 
or $v'$ is a vertex of $G_1$, including the case $v' = v$. Thus, we have 
$$ (G_1\triangleleft G_2)\triangleleft  G_3 =\sum_{v\in V(G_1)}\,\, \sum_{v'\in V(G_2)} \,\,  (G_1 \triangleleft_v G_2) \triangleleft_{v'} G_3 +
\sum_{v, v'\in V(G_1)} \,\, (G_1\triangleleft_v G_2) \triangleleft_{v'} G_3. $$
Similarly, we have
$$ G_1\triangleleft (G_2 \triangleleft G_3) = \sum_{v\in V(G_1)} \sum_{v'\in V(G_2)} \,\,  
G_1\triangleleft_{v} (G_2 \triangleleft_{v'} G_3), $$
where $v$ is glued to is the base vertex $v_2$ of $G_2 \triangleleft_{v'} G_3$, which is the same as
the base vertex of $G_2$. 
Thus, we obtain
$$ (G_1\triangleleft G_2) \triangleleft G_3 - G_1\triangleleft (G_2 \triangleleft G_3) 
= \sum_{v, v' \in V(G_1)} G_1\cup_{v\equiv v_2} G_2 \cup_{v' \equiv v_3} G_3 $$
and similarly
$$ (G_1\triangleleft G_3)\triangleleft G_2 - G_1\triangleleft (G_3 \triangleleft G_2) 
= \sum_{v, v' \in V(G_1)} G_1\cup_{v'\equiv v_3} G_3 \cup_{v\equiv v_2} G_2, $$
which proves \eqref{prelie}.
\endproof

We then obtain the associated Lie algebra.

\begin{cor}\label{Lialg1cor}
Let $\cG$ be a context-free Insertion Graph Grammar of rooted directed acyclic graphs, 
with start graph $G_S=\{ v\}$ and production rules $P(v,v_2,G_2)$, with $v_2$ the
source of $G_2$. Then there is an associated Lie algebra ${\rm Lie}_\cG$ given by the vector space
$\cV$ spanned by the graphs of $\cW_\cG$ with the Lie bracket $[G_1,G_2]=G_1\triangleleft G_2-
G_2 \triangleleft G_1$.
\end{cor}

\medskip

\begin{rem}\label{remsink}{\rm
A variant of the above construction that makes it (very mildly) context sensitive is obtained by
requiring that the marked source vertex of the graph $G_2$ in a production rule 
$P(v,v_2,G_2)$ is glued to a {\em sink} vertex of the graph $G_1$, to which the rule is applied.
The argument is exactly as before, and one obtains a pre-Lie insertion operator and
a Lie algebra. We will generalize this context-sensitive version to more general gluing data
in Propositions \ref{preLie2ins} and \ref{preLie3ins} below.}
\end{rem}

\medskip
\subsection{Some Lie algebras of context-sensitive grammars of directed graphs}

We now consider a variant of the case of Proposition \ref{preLie1ins} where
we consider an example of context-sensitive graph grammars. 
We still assume, as above,
that $\cG$ is an Insertion Graph Grammar $\cG$ as in Definition \ref{IGG1},
with start graph $G_S$ a single vertex, and where all the graphs $G\in \cW_\cG$
are directed. We no longer require that they are acyclic, hence graphs will 
generally have oriented loops. An oriented loop $\gamma$ in a graph $G$ 
is an attractor if all the edges in $E_{G}(\gamma)$ (defined as in \eqref{EGGprime}) are incoming,
that is, $\partial(e) \cap V(\gamma)=t(e)$. It is a repeller if all edges in $E_{G}(\gamma)$
are outgoing, $\partial(e) \cap V(\gamma)=s(e)$. In general, there will be also oriented loops 
that are neither attractors not repellerts. We modify the previous context-free construction
by considering, in addition to the production rules that glue a vertex of one graph 
to a source vertex of another, also context-sensitive production rules that glue an attractor 
loop of the first graph to a repeller loop of the second, 
\begin{equation}\label{Ploop}
G_1\triangleleft_\gamma G_2 := P(\gamma, \gamma_2, G_2)(G_1)=G_1 \cup_{\gamma\equiv \gamma_2} G_2,
\end{equation}
where the two graphs are glued by identifying the two oriented loops $\gamma$ and $\gamma_2$
(which necessarily have to have the same number of edges). The insertion operator is then
defined as
\begin{equation}\label{insopH}
G_1 \triangleleft G_2 = \sum_{\substack{\gamma \subset G_1 \\ \gamma \text{ attractor loop} 
\\ \gamma \simeq \gamma_2}} G_1 \triangleleft_\gamma G_2 =\sum_{\substack{\gamma \subset G_1 \\ \gamma \text{ attractor loop}\\ \gamma \simeq \gamma_2}} 
P(\gamma, \gamma_2, G_2)(G_1).
\end{equation}

\begin{prop}\label{preLie2ins}
Given a context-sensitive Insertion Graph Grammar $\cG$ as above,
the insertion operator \eqref{insopCF} defines a pre-Lie
structure on the vector space $\cV$ spanned by the graphs in $\cW_\cG$.
\end{prop}

\proof The composition $G_1 \triangleleft (G_2 \triangleleft G_3)$ is given by
$$ G_1 \triangleleft (G_2 \triangleleft G_3) = 
\sum_{\gamma \subset G_1} \,\, \sum_{\gamma' \subset G_2}
G_1 \triangleleft_\gamma (G_2 \triangleleft_{\gamma'} G_3) $$
while the composition $(G_1 \triangleleft G_2) \triangleleft G_3$ is
$$ (G_1 \triangleleft G_2) \triangleleft G_3 = 
\sum_{\gamma \subset \Gamma_1} \,\, \sum_{\gamma'\subset  G_1 \triangleleft_\gamma G_2}
(G_1 \triangleleft_\gamma G_2) \triangleleft_{\gamma'} G_3. $$
In the last sum, the choice of $\gamma' \subset G_1 \triangleleft_\gamma G_2$ can be
broken down into the case where $\gamma' \subset G_1$, 
the case where $\gamma' \subset G_2\smallsetminus \gamma$, and the case where it intersects
both, $\gamma' \cap G_1\neq \emptyset$ and $\gamma' \cap G_2\smallsetminus \gamma\neq \emptyset$.
In fact, because of our assumptions on the production rules, only the first two
possibilities can occur, and the first one can occur only with $\gamma'\cap \gamma =\emptyset$
To see this, suppose $\gamma'$ intersects both sets. Then it must intersect $\gamma$, since
$\gamma'$ is connected and $\gamma$ is the frontier between $G_1$ and $G_2$.
Under our assumptions, $\gamma'$ is an attractor loop for $G_1\triangleleft_\gamma G_2$,
hence all edges in $E_{G_1\triangleleft_\gamma G_2}(\gamma')$ must be incoming to $\gamma'$.
On the other hand, $\gamma$ is an attractor loop for $G_1$ and a repeller loop for $G_2$,
so inside $G_1\triangleleft_\gamma G_2$, there are vertices of $\gamma$ that have both incoming
and outgoing edges in $E_{G_1\triangleleft_\gamma G_2}(\gamma)$. 
Consider a vertex $v$ in the intersection $\gamma'\cap \gamma$. Either $\gamma'$ and $\gamma$
have an adjacent edge in common, or they cross each other transversely at $v$. If they are
transverse, then the incoming and outgoing edges of $\gamma$ at $v$ show that $\gamma'$
cannot be an attractor loop for $G_1\triangleleft_\gamma G_2$. If $\gamma$ and $\gamma'$ have
at least one edge adjacent to $v$ in common, then that edge is either incoming or outgoing 
at $v$. If it is incoming, then the next edge of $\gamma$ is outgoing and that suffices
to show $\gamma'$ is not an attractive loop. If it is outgoing, then one can argue the same
way with the next vertex. Thus, we can rewrite the sum above as
$$ (G_1 \triangleleft G_2) \triangleleft G_3 = 
\sum_{\gamma \subset \Gamma_1} \,\, \sum_{\gamma'\subset  G_1 \smallsetminus \gamma}
(G_1 \triangleleft_\gamma G_2) \triangleleft_{\gamma'} G_3 +
\sum_{\gamma \subset \Gamma_1} \,\, \sum_{\gamma'\subset G_2\smallsetminus \gamma}
(G_1 \triangleleft_\gamma G_2) \triangleleft_{\gamma'} G_3. $$ 
Notice that, in the sum describing $G_1 \triangleleft (G_2 \triangleleft G_3)$
we also have $\gamma' \cap \gamma =\emptyset$ because $\gamma' \subset G_2\triangleleft_{\gamma'} G_3$
is not a repelling loop so it cannot intersect the repelling loop $\gamma_2$ that is glued to $\gamma$.
We then obtain
$$ (G_1\triangleleft G_2) \triangleleft G_3 - G_1\triangleleft (G_2 \triangleleft G_3) 
= \sum_{\substack{\gamma, \gamma' \subset G_1 \\ \gamma,\gamma' \text{ attractor loops}}} 
G_1\cup_{\gamma\equiv \gamma_2} G_2 \cup_{\gamma' \equiv \gamma_3} G_3 $$
and similarly
$$ (G_1\triangleleft G_3)\triangleleft G_2 - G_1\triangleleft (G_3 \triangleleft G_2) 
= \sum_{\substack{\gamma, \gamma' \subset G_1 \\ \gamma,\gamma' \text{ attractor loops}}} 
G_1\cup_{\gamma'\equiv \gamma_3} G_3 \cup_{\gamma \equiv \gamma_2} G_2, $$
which proves \eqref{prelie}.
\endproof

We obtain an associated Lie algebra ${\rm Lie}_\cG$, as in Corollary \ref{Lialg1cor}. This
construction can be further generalized to other context-sensitive grammars, in the following way.
Assume again that $\cG$ is an Insertion Graph Grammar $\cG$ as in Definition \ref{IGG1},
with start graph $G_S$ a single vertex, and where all the graphs are oriented. 
In addition to the production rules that
glue a vertex of one graph to a source vertex of another, 
as in Proposition \ref{preLie1ins}, we also allow for
context-sensitive production rules of the form $P(G_L,H,G_R)$ where $H$ is
a connected, oriented graph with no sources or sinks. Moreover, we require that all the
edges in $E_{G_L}(H)$ are incoming to $H$ and all the edges in $E_{G_R}(H)$
are outgoing from $H$. We define the insertion operator by setting
\begin{equation}\label{GLHins}
 G_1\triangleleft_{G_L,H} G_2:= P(G_L,H,G_2)(G_1) = G_1\cup_H G_2 
\end{equation} 
to be the gluing of $G_1$ and $G_2$ along $H$, whenever $G_1$ contains a pair of subgraphs
isomorphic to $H\subset G_L$, with the orientation requirements as specified above.
We then have
\begin{equation}\label{GLHinsop}
 G_1\triangleleft G_2:=\sum_{G_L \subset G_1} P(G_L,H,G_2)(G_1),  
\end{equation}
where the sum is over all the production rules and over all the possible 
ways of identifying $G_L$ with a subgraph of $G_1$.
The result is zero if $G_1$ does not contain any subgraph isomorphic to $G_L$.
We then have the following straightforward generalization of Proposition \ref{preLie2ins}.

\begin{prop}\label{preLie3ins}
Given a context-sensitive Insertion Graph Grammar $\cG$ as above,
the insertion operator \eqref{GLHinsop} defines a pre-Lie
structure on the vector space $\cV$ spanned by the graphs in $\cW_\cG$.
\end{prop}

\proof The argument follows along the same lines as the previous cases.
Observe that, since $H$ is a directed graphs with neither sinks nor sources,
at every vertex $v\in V(H)$ there are at least one incoming and one outgoing
edge in $E(H)$. We then argue exactly as in Proposition \ref{preLie2ins}.
In the composition 
$$ (G_1 \triangleleft G_2) \triangleleft G_3 = 
\sum_{H\subset G_L \subset \Gamma_1} \,\, \sum_{H'\subset G_L' \subset  G_1 \triangleleft_{G_L,H} G_2}
(G_1 \triangleleft_{G_L,H} G_2) \triangleleft_{G_L',H'} G_3, $$
if the subgraph $H' \subset G_1 \triangleleft_{G_L,H} G_2$ has nontrivial intersection
with both $G_1$ and $G_2\smallsetminus H$, then it must intersect $H$. Then the
conditions on the orientations imply that $H'$ cannot have only incoming edges in
$E_{G_1 \triangleleft_{G_L,H} G_2}(H')$, which contradicts the orientation 
requirements for $H'$. So the only compositions that give non-trivial terms are
the ones where $H'$ is fully contained in either $G_1$ (in fact $G_1\smallsetminus H$)
or in $G_2\smallsetminus H$. We then write the composition above by separating out
the sums for these two cases, and the rest of the argument follows exactly as in the
previous proposition.
\endproof

\medskip
\subsection{Lie algebras and gluing along half-edges} 
We now consider Insertion Graph Grammars $\cG$ as in Definition \ref{IGG2}
and we consider the case where the left-hand-side of the production rules are
corollas $C(v)$, though we do not require that they are all of the same valence, so that
we include grammars that are (mildly) context-sensitive. Graphs are not
necessarily oriented, but we assume that they have a base vertex.
For a graph $G$ and a vertex $v\in V(G)$ we write $E_{ext}(G,v) \subset E_{ext}(G)$ 
for the subset of external edges of $G$ that are attached to the vertex $v$.
The condition on external
edges in Definition \ref{IGG2} corresponds in this case to the requirement that
the number of external edges $E_{ext}(G_2,v_2)$ attached to the base vertex
$v_2$ is at least equal to the valence of $C(v)$. 
These external edges of $G_2$ are identified by the production rule with
the half-edges of the corolla $C(v) \subset G_1$, where $G_1$ is the
graph the production is applied to. The argument is then exactly as in 
Proposition \ref{preLie1ins}.  The base vertex of a production $P(C(v), C(v_2), G_2)(G_1)$
is the base vertex $v_1$ of $G_1$

\begin{prop}\label{preLie4ins}
Let $\cG$ be an Insertion Graph Grammars $\cG$ as in Definition \ref{IGG2}, where
all the production rules are of the form $P(C(v),C(v_2),G_2)$ with $\val(C(v))=\val(C(v_2))$
and $E_{ext}(C(v))\subset E_{ext}(G_2,v_2)$. Assume the start graph $G_S$ of $\cG$ is also a
corolla $C(v)$. The insertion operator 
\begin{equation}\label{insopCF2}
G_1\triangleleft G_2 = \sum_{\substack{v \in V(G_1) \\ {\rm val}(v)={\rm val}(C(v_2))}} 
P(C(v),C(v_2),G_2)(G_1) =\sum_{\substack{v \in V(G_1) \\ {\rm val}(v)={\rm val}(C(v_2))}}
G_1\triangleleft_{C(v)} G_2
\end{equation}
defines a pre-Lie structure on the vector space $\cV$ generated by the graphs in $\cW_\cG$.
\end{prop}

\proof We have 
$$ G_1 \triangleleft (G_2 \triangleleft G_3) = 
\sum_{\substack{v \subset V(G_1) \\ \val(v)\leq \# E_{ext}(G_2\triangleleft_{C(v')} G_3,v_2)}} \,\, 
\sum_{\substack{v' \subset V(G_2)\\ \val(v') \leq \# E_{ext}(G_3,v_3)}}
G_1 \triangleleft_{C(v)} (G_2 \triangleleft_{C(v')} G_3) $$
$$ (G_1 \triangleleft G_2) \triangleleft G_3 = 
\sum_{\substack{v \subset V(G_1) \\ \val(v)\leq \# E_{ext}(G_2,v_2)}} \,\, \sum_{\substack{v'\subset  V(G_1 \triangleleft_\gamma G_2)\\ \val(v')\leq \# E_{ext}(G_3,v_3) }}
(G_1 \triangleleft_{C(v)} G_2) \triangleleft_{C(v')} G_3. $$
In the first sum $E_{ext}(G_2\triangleleft_{C(v')} G_3,v_2)$ is $E_{ext}(G_2,v_2)$ since the
base vertex $v_2$ of $G_2\triangleleft_{C(v')} G_3$ is the base vertex of $G_2$. We then
separate out the last sum of the second expression into the cases where $v' \in V(G_1)$
or $v'\in V(G_2)$. We obtain, as in the previous cases, 
$$ (G_1\triangleleft G_2) \triangleleft G_3 - G_1\triangleleft (G_2 \triangleleft G_3) 
= \sum_{v, v' \in V(G_1)} G_1\cup_{C(v)} G_2 \cup_{C(v')} G_3 =
 (G_1\triangleleft G_3)\triangleleft G_2 - G_1\triangleleft (G_3 \triangleleft G_2) . $$
\endproof

Again we obtain an associated Lie algebra ${\rm Lie}_\cG$.
One can also similarly extend the case of gluing along oriented loops, by assigning orientations to the
attached half-edges, with incoming/outgoing requirements as in Proposition \ref{preLie2ins}, or in
the case of gluing along more general graphs $H$ with orientation requirements as in
Proposition \ref{preLie3ins}, reformulated in terms of half-edges. These analogs of Proposition \ref{preLie2ins}
and Proposition \ref{preLie3ins} are completely straightforward and are proved by essentially the same
argument, so we will not state them explicitly here.
Within this setting, however, one cannot further extend the construction to more general context-sensitive
cases, beyond what we have seen in Proposition \ref{preLie3ins},
because the cases where the gluing data graph $H'$ intersects both $G_1$ 
and $G_2\smallsetminus H$
in $G_1\triangleleft_H G_2$ creates terms that do not cancel in the difference
$(G_1\triangleleft G_2) \triangleleft G_3 - G_1\triangleleft (G_2 \triangleleft G_3)$ and that
are not symmetric with respect to exchanging $G_2$ and $G_3$. To see more
precisely where the difficulty lies, we can write out the expression above, as before, in
the form
$$ (G_1\triangleleft G_2) \triangleleft G_3 - G_1\triangleleft (G_2 \triangleleft G_3)=
\sum_{\substack{H\subset G_1 \\ H'\subset G_1\triangleleft G_2}} (G_1\cup_H G_2)\cup_{H'} G_3
- \sum_{\substack{H\subset G_1 \\ H'\subset G_2}} G_1\cup_H (G_2\cup_{H'} G_3). $$ 
We can separate out, in the first sum the cases where $H'$ is completely contained in $G_1$
or completely contained in $G_2$, and where it intersects both graphs. The latter case,
in general, cannot be decomposed further, because it is not necessarily true that, if a
certain graph $H'$ is the gluing data of a production rule, subgraphs $G_1\cap H'$
and $G_2\cap H'$ would also occur in production rules. Thus, the term involving
subgraphs $H'$ intersecting both $G_1$ and $G_2$ does not cancel in the difference
between $(G_1\triangleleft G_2) \triangleleft G_3$ and $G_1\triangleleft (G_2 \triangleleft G_3)$
and at the same time is not symmetric with respect to interchanging $G_2$ and $G_3$,
hence one would not obtain a pre-Lie insertion operator.

\smallskip

We conclude this section by discussing a special example, which we will return to
in our application to Feynman graphs. We denote by $G_e$ the graph consisting of
a single edge $e$, identified with the union of two half-edges $e=(f,f')$ glued together by an
involution $\cI(f)=f'$.
Let $\cG$ be an Insertion Graph Grammars $\cG$ as in Definition \ref{IGG2},
with start graph $G_S$ and with production rules:
\begin{enumerate}
\item $P(G_S,\{ f,f '\}\subset \cF_{G_S},G_e)$, where the single edge graph $G_e$ 
is glued onto a pair $(f,f')$ of external half-edges of $G_S$,
\item $P(G_S,\{ f \}\subset \cF_{G_S},G_S\cup_{f'} G_e)$, where the copy of $G_S$ in the right-hand-side
of the production is glued to the one on the left-hand-side by matching the remaining external half-edge
of $G_e$ to the half-edge $f$ to form a graph $G_S\cup_{e=(f,f')} G_S$ consisting of two copies of $G_S$
glued together along an edge. 
\end{enumerate}

\begin{prop}\label{glueextedges}
Let $\cG$ be an Insertion Graph Grammar as above. 
Then the insertion operator 
\begin{equation}\label{insopext}
G_1 \triangleleft G_2 =\sum P(G_S,\cF_{G_S},G_2)(G_1) 
\end{equation}
defines a pre-Lie structure on the vector space $\cV$ spanned by the graphs in $\cW_\cG$.
\end{prop}

\proof It suffices to notice that, by the form of the production rules, 
in the composition $(G_1 \triangleleft G_2) \triangleleft G_3$ 
the gluing of $G_3$ to $G_1 \triangleleft G_2$ happens along some of the external
half-edges of $G_1 \triangleleft G_2$. The previous gluing of $G_2$ to $G_1$, in turn,
glues some external half-edges of $G_2$ to some external half-edges of $G_1$.
Thus, the remaining half-edges of $G_1 \triangleleft G_2$ are either in $G_1\smallsetminus H$
or in $G_2\smallsetminus H$, where $H$ is the set of half-edges along which the gluing
of $G_2$ to $G_1$ happened (which are no longer external edges in $G_1 \triangleleft G_2$).
This suffices then to get the pre-Lie condition, exactly as in the cases discussed previously.
\endproof 

We will see in \S \ref{FeynSec} below that this provides a different way of constructing
of Lie algebras of Feynman graphs, which is not equivalent to the insertion
Lie algebra of \cite{CoKr2}, \cite{EFBP}. More notably, it shows that the set of Feynman
graphs of a given quantum field theory is a graph language in the sense
of the theory of formal languages.

\medskip
\subsection{Lie algebras of insertion-elimination graph grammars}

We now consider the case of insertion-elimination graph grammars, as in Definitions
\ref{IEGG1} and \ref{IEGG2}. 

Let $\cG$ be an insertion-elimination graph grammar as in Definition \ref{IEGG2}.
We assume the following hypotheses:
\begin{enumerate}
\item In all the production rules $P(G_L,H,G_R)$ the graph $G_L$ is connected and
we have $H\subset \cF_{G_L}$, a set of flags (half-edges),
with $H=E_{ext}(G_R)$
\item All the graphs in $\cW_\cG$ are oriented and in all production rules the half-edges in $H$
are incoming to both $G_L$ and $G_R$.
\end{enumerate}

\begin{prop}\label{IEGGLie}
Let $\cG$ be an insertion-elimination graph grammar satisfying the two conditions above.
Then the insertion operator
\begin{equation}\label{IEGGins}
G_1 \triangleleft G_2 = \sum_{G\subset G_1} P(G, H, G_2)(G_1) 
\end{equation}
is a pre-Lie operator.
\end{prop}

\proof With the notation $G_1\triangleleft_{G,H} G_2=P(G, H, G_2)(G_1)$, we have
$$ G_1 \triangleleft (G_2 \triangleleft G_3) = \sum_{G,G'} G_1\triangleleft_{G,H} (G_2 \triangleleft_{G',H'} G_3), $$
with $G\subset G_1$ and $H\subset \cF_{G_1}$, $H=E_{ext}(G_2\triangleleft_{G',H'} G_3)$, 
and with $G'\subset G_2$, $H'\subset \cF_{G_2}$ and $H'=E_{ext}(G_3)$. Since all the external edges
of $G_3$ are glued to flags of $G_2$, in the identification along $H'=E_{ext}(G_3)$, we have
$H=E_{ext}(G_2\triangleleft_{G',H'} G_3)= E_{ext}(G_2)$. When composing in the opposite order, we
have 
$$ (G_1 \triangleleft G_2) \triangleleft G_3 = \sum_{G,G'} (G_1\triangleleft_{G,H} G_2) \triangleleft_{G',H'} G_3, $$
with $G\subset G_1$, $H\subset \cF_{G_1}$, $H=E_{ext}(G_2)$, and with
$G'\subset G_1\triangleleft_{G,H} G_2$, $H'\subset \cF_{G_1\triangleleft_{G,H} G_2}$,
$H'=E_{ext}(G_3)$. If the graph $G'$ intersects nontrivially both $G_1\smallsetminus (G\smallsetminus H)$ 
and $G_2$, then by connectedness $H'\cap H\neq \emptyset$, but the orientation conditions on the
edges of $H$ and of $H'$ are incompatible, so $G'$ must be contained in either 
$G_1\smallsetminus G$ or in $G_2$. We then obtain
$$ (G_1 \triangleleft G_2) \triangleleft G_3 - G_1 \triangleleft (G_2 \triangleleft G_3) = \sum_{\substack{G,G'\subset G_1 \\ G\cap G' =\emptyset}} (G_1\smallsetminus ((G\smallsetminus H)\cup (G'\smallsetminus H'))) \cup_H G_2 \cup_{H'} G_3 $$
which is symmetric in exchanging $G_2$ and $G_3$.
\endproof

We obtain an associated Lie algebra ${\rm Lie}_\cG$.

\bigskip

\medskip
\subsection{The insertion Lie algebra of Quantum Field Theory and primitive graphs}

In the previous sections we have shown that one can associate to certain
classes of graph grammars an insertion Lie algebra, that behaves very
similarly to the insertion Lie algebra of Quantum Field Theory of \cite{CoKr2}, \cite{EFBP}.
It is then natural to ask whether the insertion Lie algebra of Quantum Field Theory is
itself obtained from a Graph Grammar via the same procedure discussed above.
This is {\em not} the case, because it would violate the
property that graph grammars have a {\em finite} number of production rules.
In fact,  the Lie algebra of Feynman graphs is
generated by the primitive elements of the Hopf algebra. We would like to
obtain all 1PI Feynman graphs of the theory from a graph grammar
that has a single start graph $G_S$ and a finite number of production
rules $P(G_L,H,G_R)$, in such a way that the Lie bracket $[G_1,G_2]$
of the Lie algebra of quantum field theory would agree with the
Lie bracket defined by the graph grammar,
$$ [G_1,G_2] = \sum_{G_L\subset G_1} P(G_L,H,G_2)(G_1) - \sum_{G_L\subset G_2} P(G_L,H,G_1)(G_2). $$ 
For this to be the case, we see that we would need a production rule for each insertion
of a primitive graph $G$ of the Hopf algebra of the theory into a vertex with valence equal
to the number of external edges of $G$. Since there are infinitely many primitive graphs,
this would violate the requirement that the graph grammar has only finitely many production rules.
We will see in \S \ref{FeynSec} that, despite this negative result, the
Feynman graphs of a given quantum field theory are a graph language,
obtained from a graph grammar with finitely many production rules.
These graph grammars in turn define Lie algebras, by the procedure
discussed in the previous sections, which are in general not
equivalent to the insertion Lie algebra of \cite{CoKr2}, \cite{EFBP}.

\bigskip

\section{Feynman diagrams as graph languages}\label{FeynSec}

Motivated by the insertion Lie algebra of quantum field theory, \cite{CoKr2}, \cite{EFBP},
we have shown in the previous section that, under certain conditions on the production
rules, one can associate Lie algebras to Graph Grammars.
In this section, we return to the motivating example of Feynman graphs and we show that
the Feynman graphs of certain quantum field theories are examples of graph languages.
Our generative description of Feynman graphs in terms of graph
grammars can be seen as ``reading in reverse" the procedure described in \cite{BKP}, \cite{KPKB},  
that generates all Feynman graphs starting with the vacuum bubbles (no external edges) and
progressively cutting internal edges into pairs of external half-edges.

\medskip
\subsection{The $\phi^4$ Graph Language}

We first analyze the example of the $\phi^4$-theory. 
This is the scalar quantum field theory with (Euclidean) Lagrangian
density 
\begin{equation}\label{Lagphi4}
\cL(\phi)=\frac{1}{2}(\partial\phi)^{2}+\frac{1}{2}m^{2}\, \phi^{2}+\frac{1}{4!}\lambda\,\phi^{4}.
\end{equation}
The Feynman graphs of this theory have all vertices of valence four.
More precisely, we should also include valence two vertices that correspond to
the mass and kinetic terms, but we will not mark them explicitly in the diagrams.

\smallskip

\begin{prop}\label{phi4GG}
The Feynman graphs of the $\phi^4$-theory are the elements of the graph language $\cL_\cG$
generated by a graph grammar $\cG$ as in Proposition \ref{glueextedges}, with start graph
$G_S$ given by a $4$-valent corolla, and two production rules: one of the form
$P(G_S,\{ f,f '\}\subset \cF_{G_S},G_e)$, which glues together two external edges of $G_S$
and one of the form $P(G_S,\{ f \}\subset \cF_{G_S},G_S\cup_{f'} G_e)$, which glues together
two copies of $G_S$ along an edge.
At each stage in the application of one of the production rules, the external edges of the
resulting graph are marked either with a terminal or with a non-terminal symbol. 
\end{prop}

\proof Whenever the external edges of a graph in $\cW_\cG$ are marked by non-terminal symbols
one can continue to apply production rules to them, while if all the external edges are marked by 
terminals the resulting graph is in $\cL_\cG$. Thus, a graph is in $\cL_\cG$ if either it has no more
external edges, in which case it is a vacuum bubble of the $\phi^4$-theory, or if all the external
edges are marked by terminals, in which case, it can be identified with the result of cutting a number
of edges of a vacuum bubble into half edges. This produces all Feynman graphs of the $\phi^4$
theory, \cite{KPKB}.
\endproof

\smallskip

The production procedure is illustrated in Figure \ref{phi4GG}.
The graph in the upper left-hand corner is $G_S$ for this
theory. Single arrows indicate the first type of production gluing 
external edges to make a loop. Double arrows indicate the second type
of production joining base graphs along an edge to make a new 
graph with no loops.
the graphs with no external edges are represented on the right, while
the graphs with as many external edges as possible are on the left.
Note that the latter type of graphs are trees.
t is clear that for this theory, any allowed Feynman graph
can be transformed into a tree such as the ones above. Consider the
graphs on one internal vertex. They all can be constructed by gluing
the edges of the base graph $G_S$, via repeated application of the
production rule represented by the single arrows. A similar statement 
can be made of the graphs on two internal vertices, except that, after cutting
all edges into pairs of half edges, one is now left with a disjoint union of
two copies of the base graph $G_S$. In general, any (connected) $\phi^4$ graph with 
$k$ valence $4$ vertices can be constructed by gluing together pairs of half-edges,
starting from $k$ copies of $G_S$, hence by repeated applications of the two
types of production rules.

\begin{figure}
\includegraphics[scale=0.6]{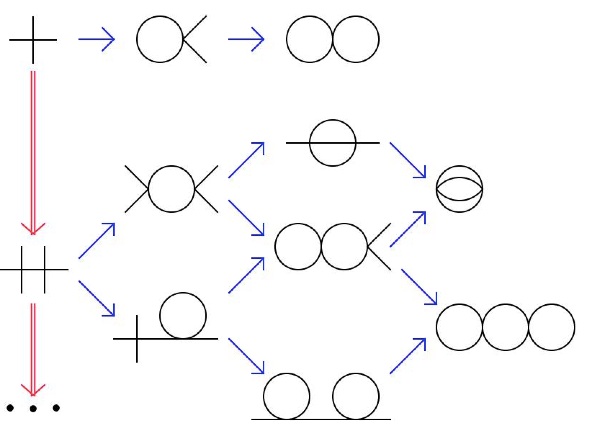}
\caption{Generative grammar for Feynman graphs: $\phi^4$-theory. \label{phi4GG}}
\end{figure}

\medskip
\subsection{Graph grammar for $\phi^k$-theories}

The case discussed above of the $\phi^4$-theory can easily be generalized to
the case of (Euclidean) Lagrangian densities
\begin{equation}\label{LagPphi}
\cL(\phi)=\frac{1}{2}(\partial\phi)^{2}+\frac{1}{2}m^{2}\, \phi^{2}+P(\phi)
\end{equation}
where the interaction term is a single monomial $P(\phi) =\frac{1}{k!} \lambda \, \phi^k$.

\begin{prop}\label{phikGG}
The Feynman graphs of the $\phi^k$-theory are the elements of the graph language $\cL_\cG$
generated by a graph grammar $\cG$ as in Proposition \ref{glueextedges}, with start graph
$G_S$ given by a $k$-valent corolla, and two production rules: one of the form
$P(G_S,\{ f,f '\}\subset \cF_{G_S},G_e)$, which glues together two external edges of $G_S$
and one of the form $P(G_S,\{ f \}\subset \cF_{G_S},G_S\cup_{f'} G_e)$, which glues together
two copies of $G_S$ along an edge.
At each stage in the application of one of the production rules, the external edges of the
resulting graph are marked either with a terminal or with a non-terminal symbol. 
\end{prop}

\proof
Let $G$ be a connected graph of the $\phi^k$ theory. Thus, all
vertices $v\in V(G)$ have valence $\val(v)=k$. We neglect for the moment 
the possible presence of valence $2$ vertices associated
to the kinetic and mass terms in $\cL(\phi)$. Consider all possible ways
of cutting internal edges, so that they are replaced by a pair of external half-edges,
that leave the graph connected. The number of possible such cuts is the
degree of edge-connectedness of the graph. Stop when no further such cuts
remain. If we denote the resulting graph by $G'$, then it is clear that $G$
is obtained from $G'$ by repeatedly applying the first production rule.
Every connected graph has a decomposition into a tree with insertions at the vertices
of 1PI graphs (one particle irreducible, also known as 2-edge-connected)
that have a number of external edges equal to the valence of the tree vertex.
By repeatedly cutting non-disconnecting edges, and using this decomposition,
it is clear that the resulting graph $G'$ is a tree. Since all the vertices of $G'$ 
have valence $\val(v)=k$, the tree can be constructed by repeated application of
the second production rule. One can consider also the presence
of valence two vertices, with each gluing of a copy of $G_e$ in the production
rules involving a valence two vertex inserted in the middle of an edge
connecting two valence $k$ vertices, and the argument remains essentially the same.
The difference between valence two vertices coming from the kinetic and the
mass terms can be taken care of by using two different terminal symbols
labeling the vertices.
\endproof

\medskip
\subsection{Graph grammars for arbitrary scalar field theories}

We then consider the case of a scalar field theory with Lagrangian density \eqref{LagPphi}
where the interaction term is a polynomial
\begin{equation}\label{Pphi}
P(\phi)=\sum_{k\geq 3}\frac{\lambda_{k}}{k!}\phi^{k}.
\end{equation}

\smallskip

\begin{prop}\label{phikGG}
The Feynman graphs of a scalar field theory with interaction polynomian $P(\phi)$ as in \eqref{Pphi} of degree $N$
are the elements of the graph language $\cL_\cG$
generated by a graph grammar $\cG$ as in Proposition \ref{glueextedges}, with start graph
$G_S$ given by a $k$-valent corolla, where $k$ is the smallest term in \eqref{Pphi}
with $\lambda_k\neq 0$ and three production rules: 
\begin{enumerate}
\item The first kind of production rules is of the form
$P(G_S,\{ f,f '\}\subset \cF_{G_S},G_e)$, which glues together two external edges of $G_S$.
\item The second kind of production rule 
$P(G_S, G_e, G_{S,f})$ glues a copy of $G_e$ to the start graph $G_S$
by identifying one half edge of $G_e$ with one of the half-edges of $G_S$ and leaving
the other half edge $f$ as a new external edge, thus creating a corolla of valence $k+1$.
If $\lambda_{k+1}\neq 0$ the vertex of the resulting graph $G_{S,f}$ can be labeled by either a terminal
or a nonterminal symbol, if $\lambda_{k+1}=0$ it is labeled by a nonterminal symbol. 
Production rules $P(G_{S,f_1,\ldots,f_r}, G_e, G_{S,1,\ldots,f_r,f'})$ can be further applied to previously produced
corollas $G_{S,f_1,\ldots,f_r}$ with vertex labeled by nonterminals, until $\val(G_{S,f_1,\ldots,f_r})=N-1$: 
in this case the vertex in the resulting $G_{S,1,\ldots,f_r,f'}$ can only be labelled by a terminal.
\item The third kind of production rules $P(G_{S,f_1,\ldots, f_r}, \{ f_i \}\subset \cF_{G_S},G_{S,f_1,\ldots, f_r}\cup_{f_i=f_j'} G_{S,f'_1,\ldots, f'_s})$, which glues together along an edge two corollas $G_{S,f_1,\ldots, f_r}$
and $G_{S,f'_1,\ldots, f'_s}$ produced by the previous type of production rules.
\end{enumerate}
At each stage in the application of one of the production rules, the external edges of the
resulting graph are marked either with a terminal or with a non-terminal symbol. 
\end{prop}

\proof The argument is similar to the previous case: one starts from an arbitrary connected
Feynman graph $G$ of the theory and performs the maximal number of cuts of internal
edges into pairs of external half-edges that leaves the graph connected. The only
difference in the argument is that the resulting graph $G'$ is now a tree with vertices of
valences ranging among the values $3\leq k \leq N$ for which $\lambda_k\neq 0$ in \eqref{Pphi}. 
These are then obtained by repeated application of the production rules of the second and
third type that produce corollas of the right valences and glue them together along edges to form the
tree $G'$.
\endproof

\smallskip

\begin{figure}
\includegraphics[scale=0.6]{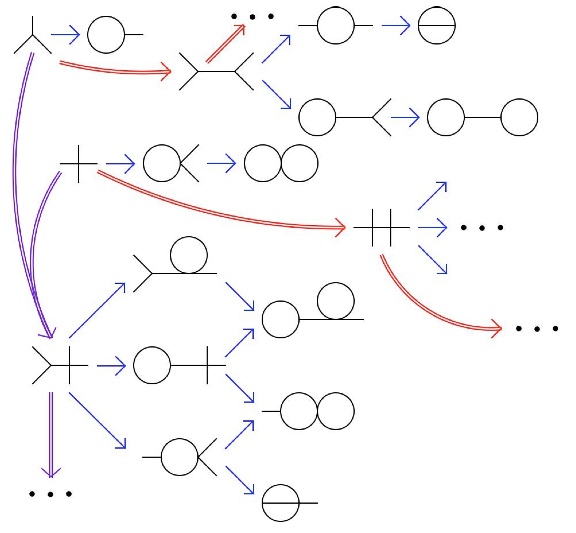}
\caption{Generative grammar for $P(\phi)=\frac{1}{6} \lambda_3\, \phi^3 +\frac{1}{24} \lambda_4\, \phi^4$. \label{phi3phi4GG}}
\end{figure}

Figure \ref{phi3phi4GG} illustrates the generative grammar of Proposition \ref{phikGG}
for the scalar field theory with 
$$ \cL(\phi) = \frac{1}{2}(\partial\phi)^{2}+\frac{1}{2}m^{2}\, \phi^{2}+\frac{1}{6} \lambda_3\, \phi^3 +\frac{1}{24} \lambda_4\, \phi^4. $$

\medskip
\subsection{Graph grammar for the $\phi^2 A$-theory}

This theory has two different propagators for the fields $\phi$ and $A$, which one
represents by drawing straight edges for the $\phi$-propagator and wavy edges
for the $A$-propagator. The cubic interaction terms implies that the Feynman graphs
have vertices of valence $3$ with two straight and one wavy external half-edges. 
The Feynman graphs of this theory were analyzed in \cite{KPKB}, with $\phi$
representing fermions and $A$ the photon. Note that in this theory 
edges representing photons are always internal, unlike what happens in
quantum electrodynamics, where photons can be external, \cite{BKP}. 

\begin{prop}\label{phi2AGG}
The Feynman graphs of the $\phi^2 A$-theory are the elements of the graph language $\cL_\cG$
generated by a graph grammar $\cG$ as in Proposition \ref{glueextedges}. The start graph
$G_S$ has two trivalent vertices, one internal wavy edge connecting them, and four external
straight half-edges. There are two kinds of production rules: one of the form
$P(G_S,\{ f,f '\}\subset \cF_{G_S},G_e)$, which glues together two external edges of $G_S$
and one of the form $P(G_S,\{ f \}\subset \cF_{G_S},G_S\cup_{f'} G_e)$, which glues together
two copies of $G_S$ along an external edge.
At each stage in the application of one of the production rules, the external edges of the
resulting graph are marked either with a terminal or with a non-terminal symbol. 
\end{prop}

\proof The graph grammar $\cG$ is illustrated in Figure \ref{phi2AGG}. 
Notice that, if photon edges were allowed to be external, then the argument would
be the same as in the $\phi^{3}$-theory, except that the labeling of the edges
as bosonic or fermionic must be taken into account when inserting the base graph 
in the tree. The fact that we require bosonic edges to be internal means that these
edges cannot be cut in the process that leads from $G$ to $G'$.
SInce all vertices in $G$ have valence $3$
with two fermion and one boson line, after all the internal fermion lines
are cut, one still obtains a tree $G'$, which we now view as
being formed out of repeated application of the second production
rule applied to the start graph.
\endproof

The properties of external and internal edges of the $\phi^2 A$-theory
discussed in \cite{KPKB} are reflected here in the fact that, in the
production rules, it is only possible to join
base graphs along fermion edges. The fact that photon
edges are only internal is taken into account by the choice
of the start graph having two vertices instead of one, with one
internal bosonic edge.

\begin{figure}
\includegraphics[scale=0.5]{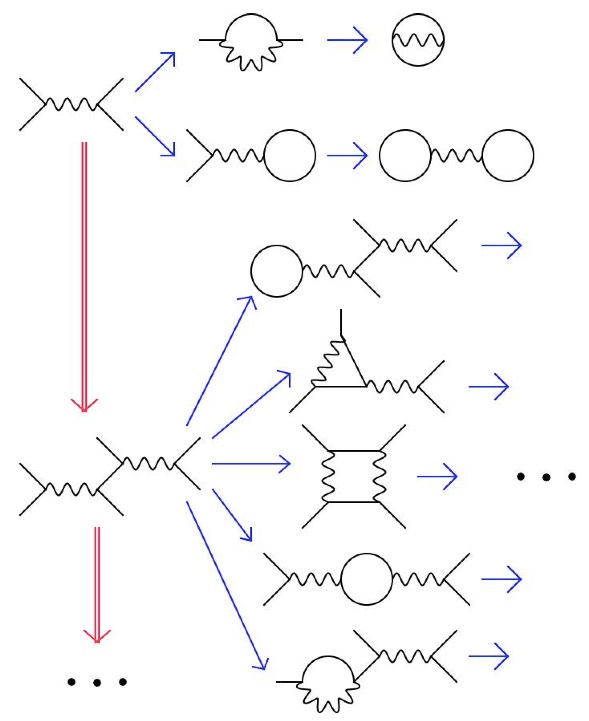}
\caption{Generative grammar for the $\phi^2A$-theory. \label{phi2AGG}}
\end{figure}

\medskip
\subsection{A general procedure}

We can summarize all the cases discussed above, for the different theories, in
a common general procedure, as follows.

\begin{itemize}

\item Fix $n\in\mathbb{Z}^{+}$. This is the number of distinct conditions.
\item Fix $t\in\mathbb{Z}^{+}$. This is the number of edge types. Let $v\in\{0,1\}^{t}$
where $v(i)$ is 0 if type $i$ edges cannot be external and 1 otherwise.
\item For each $k\in\{1,...,n\}$:
\begin{itemize}
\item Let $C_{k}\in M_{t,2}(\mathbb{Z})$ where $C_{k}(i,1)=i$ for all
$i\in\{1,...,t\}$ (the edge type) and with $C_{k}(i,2)$ the number
of edges of type $i$ allowed by this condition at each vertex.
\item Let $G_{k}$ be the star graph with edges determined by $C_{k}$
\end{itemize}
\item Let $G$ be a connected graph that satisfies the conditions of $v$
and each of the $C_{k}$. 
\item There is a sequence of production rules that glue
edges of finitely many copies of the graphs $G_{k}$ to make $G$.
\item Any $G$ that satisfies these conditions can be constructed
from these initial graphs using the production rules.
\item In order to have a single start graph one needs to add further
production rules that derive higher valence star graphs $G_{k}$
from lower valence ones, marking the vertex with a terminal label
when the process should stop.
\end{itemize}

\begin{ex}\label{ex1}{\rm 
In a $\phi^{2}A$ theory, $n=1$, $t=2$, $v=(1,0)$
and 
$C_{1}=\left(\begin{array}{cc}
1 & 2\\
2 & 1
\end{array}\right)$.}
\end{ex}

\begin{ex}\label{ex2}{\rm 
In a theory where $P(\phi)=\frac{\lambda_{3}}{3!}\phi^{3}+\frac{\lambda_{4}}{4!}\phi^{4}$,
$n=2$, $t=1$, $v=(1)$, $C_{1}=(\begin{array}[t]{cc}
1 & 3\end{array})$ and $C_{2}=(\begin{array}[t]{cc}
1 & 4\end{array})$.}
\end{ex}

\bigskip
\bigskip

\noindent {\bf Acknowledgment} The first author is supported by NSF grants DMS-1007207, 
DMS-1201512, PHY-1205440. The second author was supported by a Summer Undergraduate 
Research Fellowship at Caltech.

\end{document}